# Adaptive Zone-2 Distance Protection Scheme for Multi-Terminal Line Connecting Wind Farm

Seyede Fatemeh Hajeforosh [a1], Nabiollah Ramezani [b], Ali AhmadiDounchali [c]

[a1] Department of Electric Power Engineering, Lulea University of Technology, Lulea, Sweden

[b] Department of Electrical and Computer Engineering, University of Science and Technology of Mazandaran, Behshahr, Iran

[c] Department of Electrical and Computer Engineering, Qazvin Azad University, Qazvin, Iran

**Abstract**

Wind energy is one of the fastest growing renewable sources in the world that converts wind power into the electricity through wind farms. Stochastic nature of the wind power makes interconnection of wind farms to the grid unreliable. Thus, in order to enhance the system stability, it is important to protect different parts of the power system specifically transmission lines. Since distance relays are the primary devices in the protection of lines, adjusting their setting is a crucial issue in terms of mitigating the rate of disruption and increasing the overall stability. In this paper, the impact of wind speed variation in the second zone of the distance relay has been studied. To monitor distance relay operation part of a power system is simulated in PSCAD/EMTDC. Then, an adaptive Technique is proposed to coordinate zone-2 operation in accordance to other zones and prevent over/under reach. This method is based on the Neural Network in MATLAB Simulink to adjust the second zone automatically due to occurring instantaneous changes in the grid.

## 1    Introduction

Since many years ago transmission lines have been utilized to deliver power electricity from generation units to consumers [1]. Transferring power with minimum losses, high efficiency, and stability needs reliable protecting scheme across the network especially transmission lines [2-3]. One of the key components regarding such protection is distance relays which are widely used as a primary and backup protection [4-5]. However, in recent years introducing innovative technologies such as wind energy has brought significant changes in the performance of protective devices [6].

Wind farms are increasingly integrated to power grids at different voltage levels all over the world [7-9]. The share of such farms in a power system causes fundamental problems that may lead to system uncertainty [10]. Clearly, wind speed fluctuation is one of those destructive factors that causes cascading failures [11-12]. Due to the non-linear relationship between output power of each generation units and wind speed; once the speed exceeds the limits, only a few of wind turbines remain in service while others stop working. Thus, the transmission system connected into wind farms will be exposed to the permanent variable condition [13]. In such cases, distance relays cannot detect faults from transient disturbances accurately and it may result in their maloperation and false trip. Due to the significant role of distance relay in interconnected networks, in recent years a lot of work has been carried out to adjust relay setting;    however, most of them focused on the operation of wind farm side relay and presented adaptive setting based on numerical analysis [14-16]. In [17] an adaptive scheme is proposed based on quadrilateral relay characteristic that sets the trip boundary only with local information derived from wind

---

farm side. This is an effective method for the low penetration of wind farm; however, for a higher level of wind capacity, the information from the main grid is needed. Following [17] scheme, in [18] authors propose a separate adaptive unite in association with default setting unit by using local information from the wind farm. The characteristic of the unit is constructed with output points of seven neural networks leading to the accurate operation in adapting to the instant conditions of the network. Studying the protection of wind power distributed generation is the focus of [19]. It is based on the analysis of distance relay dynamic behavior by using pre-fault voltages for mho relay types. The main concept behind this method is to increase the coverage of fault resistance as well as detecting hidden failures behind the relay. Discussing the operation of distance relays in identifying internal faults among wind turbines is the purpose of research works in [20-21]. The protection schemes rely on estimating the impedance at the common coupling point (PCC) in order to extend the zone-1 reach to nearly cover the whole line up to the next adjacent bus.

The aforementioned methods may be favorable for an ideal trip characteristic of wind farm side distance relay in case of steady-state modeling and fixed wind speed. However, if the wind varies considerably, precisely in radial networks including several in-feeds; fault conditions may cause a destructive impact on the reliability of distance relays. To reduce malfunction, in this paper the effect of wind speed variation on the performance of grid side zone-2 distance relay has been taken in to account. Moreover, an effective adaptive algorithm is proposed to implement and provides accurate settings due to the prevailing system conditions. Results are provided for single phase to ground fault and the concept can be extended to other types of faults as well.

The rest of the paper is organized as followed. Section 2, introduces the distance relays and their setting in transmission lines. In Section 3 the impact of wind speed in a part of the power system is being addressed. Section 4 is dedicated to proposing a new adaptive method for zone-2 adjustment and finally, the conclusion is depicted in section 5.

## 2 Distance protection

Distance relays are normally used to protect transmission lines from different types of faults. They normally consist of three zones with coordinated delays [22]. The first zone detects faults on 80%-90% of the protected line without any intentional delay. The second and third zones are usually operated as backup protection for part of the neighboring line with an appropriate delay of 0.3s and 1s respectively. Besides, the second zone is set to reach nearly 50% of the next adjacent line while the reach of the third zone is set up to 80% of the next line section [23]. The configuration of the conventional transmission system protecting by different zones of distance relay is shown in Fig. 1.

Due to the importance of calculating exact impedance seen by distance relays in transmission lines especially in case of any failure, it is crucial to provide reliable back up protection to prevent any unintentional trip. The second zone of such relays is the immediate protection if zone 1 fails to operate normally by protecting adjacent line without overlapping. Calculating zone-2 setting is the focus of this part that has been carried out in two states; ordinary transmission lines and multi-terminal lines. According to Fig.1 the impedance setting of zone-2 relay provided at bus A is provided in equation (1):

$$Z_{set2p} = Z_{AB} + 0.5Z_{BC} \qquad (1)$$

In this equation $Z_{set2p}$ is the impedance setting for second zone of relay AB, $Z_{AB}$ is defined as an impedance of the protected line A-B and $Z_{BCi}$ an impedance of the shortest remote line B-$C_i$. In the above formula, i is equal to 1, 2... K in which K is the total number of primary relays that are applied to the part of a power system.

### 2.1 Zone-2 setting in multi-terminal lines

The ideal transmission lines are two terminals with no additional loads at remote ends; however, in reality, it is hardly possible to achieve such system since most networks involve multiple lines with tapped loads [24]. Due to the nature of transmission lines with multiple generating units, several problems occur in the protective device's performance especially distance relays. Increasing the number of generating sections means more current injection into the grid causing maloperation of relays. Two major problems regarding multi-terminal lines protection are overreaching and underreaching in which they are defined as situations when the relay commands trip for faults beyond its setting and operates for faults within the protected zones respectively [25]. In such cases in order to prevent abnormal errors, some changes are added to the zones setting to increase selectivity and security of the relays. The scheme of multiple lines with distance relays is depicted in Fig.2.

In order to determine the zone-2 relay seen impedance, firstly we should determine the voltage at bus A, consider equation (2).

$$E_A = Z_A I_{Relay} + Z_f(I_{Relay} + I_{Remote})$$ (2)

$E_A$ is the voltage at bus A, $Z_{Relay}$ and $Z_f$ are grid and fault side's impedance respectively.

$I_{Remote}$ is the current flowing from generating units while $I_{Relay}$ is the flowing current from the relay side.

Then the next step is obtaining the impedance setting according to injected currents [26] provided in equation (3). Note that, in radial networks with multiple in feeds the second zone is set to reach the neighboring line with the lowest positive sequence impedance:

$$Z_{set2s} = Z_{AB} + 0.5 \ MinZ_{BC_i}\left(1 + \frac{I_{Remote}}{I_{Relay}}\right)$$ (3)

In this equation the parameters are as followed:

$Z_{set2s}$: Impedance setting for second zone of relay AB

$Z_{AB}$: Impedance of the protected line A-B

$Z_{BCi}$: Impedance of the shortest remote line B-$C_i$

$i$: Equal to 1, 2… K and K is the total number of primary relays

In order to facilitate further studying $K_{Remote}$ is defined [27] as the fraction of the remote lines protected by the zone-2 relay and calculated through equation (4).

$$K_{Remote} = 1 + \frac{I_{Remote}}{I_{Relay}}$$ (4)

$K_{Remote}$ is the determiner factor in achieving a reliable setting; however in most research works it is considered as a constant parameter by line coverage percentage. Hence, in the following section we want to study the impact of $K_{Remote}$ as an influential coefficient on multiple transmission lines connecting a wind farm. Due to the intermittent behavior of the wind speed the output voltage and current vary during the day. As a result, investigating the dynamic model of wind farm helps to demonstrate a model close to the reality with high flexibility.

## 3 Effect of Wind Power Variation on Relay Operation

Study the impact of speed variation on part of the power system is the focus of this section. To see the effect of phase to ground fault at the end point of zone-2 reach and also the capability of the conventional distance protection in detecting the fault, all the changes have been applied on a following sample network.

### 3.1 Part of the System under Consideration

Fig. 3 illustrates the system under study. The wind farm interconnection with the transmission line includes following sub-systems at 60Hz:

A 15 MW wind farm with 3MW permanent magnet synchronous generator wind turbine.

Collector substations (33kV) to connect all wind turbines.

Point of Common Coupling (PCC) substation (33kV) to connect collector substations with transmission lines.

A 132 KV Transmission line with positive sequence $30\ e^{22j}$ and zero sequence $82e^{0.5}$
The grid consists of 100 MVA synchronous machine connecting to the 132 KV bus.
One of the requirements in this study is having a dynamic model of some components in the simulation. Modeling the 15 MW wind farm with penetration levels from 0% to 100% in which 100% shows the full capacity and 0% means there is no generation from the wind farm to the line (primary adjustment). To build a wind farm, we aggregate 5 wind turbines with 3 MW. The assumption is that turbines are working in a specific range of wind speed between $4m/s$(Minimum) and $25m/s$ (Maximum). Any small changes in this range needs to be addressed as part of the study for further data-driven. Typical distance relay is modeled through Fast Fourier Transform for initiate simulation and discovering intermittent parameters.

*3.2 Simulation Results*

In this section, the aim is to investigate and discuss the impact of wind power generation on the performance of zone-2 Distance Relay (R1) under a certain type of fault within a different range of wind speed. The study is carried out in PSCAD/EMTDC Simulink and the results are presented.

In order to run a simulation, grid side relay zones are adjusted considering no generation from the wind farm. Since the purpose of this study is setting the trip boundary and finding out the relay operation during the fault at the end of the second zone reach point, it is assumed that each time a specific fault occurs in one of the remote lines at the nearest point to the third zone. Fig.4 shows seen impedance when the fault happens at 16km from line B.

Clearly, before applying remote power supply and posing the network towards wind fluctuations the relay operates normally according to its default settings.

However when the wind farm gets connected to the line, relay characteristic changes as a function of wind with the initiation rate of 4m∕s. Fig.5 indicates the impact of speed variations on the distance operation causing over reach. Thus, the relay detects the fault earlier than it was and may lead to a false trip.

Since during the day system may exposed to instant climate changes and these continuous fluctuations have caused system's deficiency, it is necessary to take in to account every probable variation in the operational range of wind farm. This would be done through online monitoring and measuring data. In the following, some cases are simulated on the mho characteristic of the zone-2 boundary to clarify the problem.

In Fig 6, applying $16^{m}/_{s}$ and $22^{m}/_{s}$ wind speed to the system lead to relay under reach so the command trip would be sent sooner and relay operates wrongly. In contrast to the previous cases in Fig 7 the $23^{m}/_{s}$ and $16^{m}/_{s}$ wind speed cause overreach in relay operation and thus the system's downtime. Clearly, each time mho characteristic would be in contradiction to its previous status in which alters between the second and third zones.

Implementing simulations with all different range of wind speed and a collection of data derived from measuring voltage and current in the PSCAD/EMTDC Simulink proved that the unreliable function of speed variation on the second zone of the grid distance relay in transmission lines makes the overall system unstable. Since relay should detect faults accurately in less than a fraction of a second and separates defective part from the rest, it would be necessary to see the impedance due to the prevailing operating conditions of the power system. Therefore, in the following, a practical adaptive scheme is proposed to address this issue.

**4 Adaptive Protection**

It is derived from simulation (section 3.2) that integrating wind farm to the power system has brought major changes in power grids infrastructures. Thus, Development of distance protection needs to be enhanced in accordance with these changes in order to prevent any failures or unexpected interruptions. In recent years adaptive solutions based on online monitoring have been introduced rather than conventional methods. The principle is using various techniques in which each of them focuses on one part of the system to improve overall flexibility [28-30]. In the following section, a new technique based on Neural Network Algorithm is proposed to adjust second zone distance relay by shrink - expand method.

*4.1 Proposed Adaptive Scheme*

The Neural network is one of the methods for optimizing the network [31] based on feeding some input variables to produce outputs. One of the simplest and efficient suggested layouts for modeling the neural network is Multi-Layer Perceptron (MLP) Fig.8 [32]. It is a Feed-Forward Artificial Neural Network model showing sets of input data on to sets of appropriate outputs. The MLP consists of multiple layers in which each layer is fully connected to the next one. Besides, it utilizes a supervised learning technique called Back Propagation for training the network. MLP is a modification of the standard linear Perceptron and can distinguish those data that are not linearly separable.

In this paper, a 1_input 1_output Feed-Forward Back Propagation with 85 neurons in hidden layers and a linear fitting function for the output is used as shown in Fig 9.

Distance Relay is trained to respond quickly to the speed changes. Approximately about 60%-70% of all data are randomly chosen and trained while remained parameters, about 30%-40%, are considered as an input. The accuracy of the trained network in this study is verified at Fig.10.

The above curve demonstrates the validation of trained parameters due to the close correlation between instructed values and real calculation. Therefore, it would be capable of covering instant conditions of wind fluctuations in power system. In Fig. 11, different steps of applying the proposed adaptive scheme to the relay setting are indicated.

Commonly, distance relays are set based on offline information, voltage, and current, derived from the network. When wind farm, as a power supply, connected to one of the remote lines more current is flowing through the line so it may causes the relay fails to operate or even mal operate due to the overreach or under reach. In this case, it is necessary to estimate the rate of currents emanating from remote side and relay (R1) to calculate $K_{Remote}$ and seen impedance using equation (3). When wind farm starts working relay system needs instantaneous data from that side to adjust the zones based on new information and if necessary expands or shrinks the trip boundaries. To achieve this aim we propose utilizing Neural Network Algorithm in Matlab Simulink to modify each seen impedance simultaneously. This scheme would also be applicable if the wind farm stops working by calculating the impedance through its default calculation.

# 5    Conclusion

This paper proposes a distance protection algorithm for the second zone of the distance relay in multiple transmission line connecting wind farm. The algorithm is based on wind speed variation throughout the day and calculates the apparent impedance seen by the grid side relay based on the magnitude of the voltage and current flowing from wind farm to the grid side. The simulation results demonstrate that integration of wind farm to the AC grid results the false trip for distance relays in which this method can successfully change zone-2 trip boundary instantly due to the prevailing condition of a power system. In this study, fault resistance is considered as a constant parameter while wind speed is dynamic. Thus, further studying can be conducted considering both factors as variables and extend the simulation for different types of faults. Then the proposed adaptive protection can be applied to the real networks to verify its accuracy.

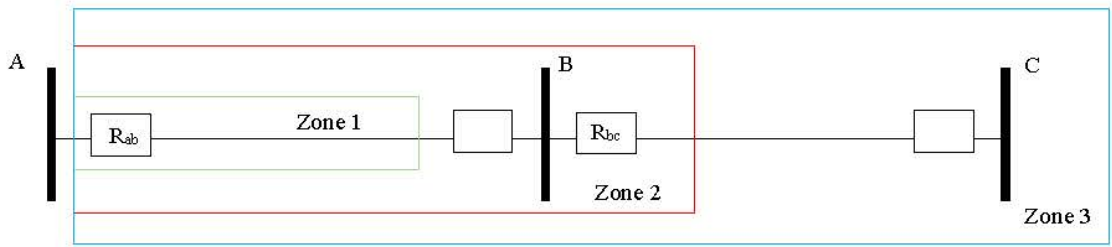

**Fig.1.** Traditional transmission system protected by distance relays.

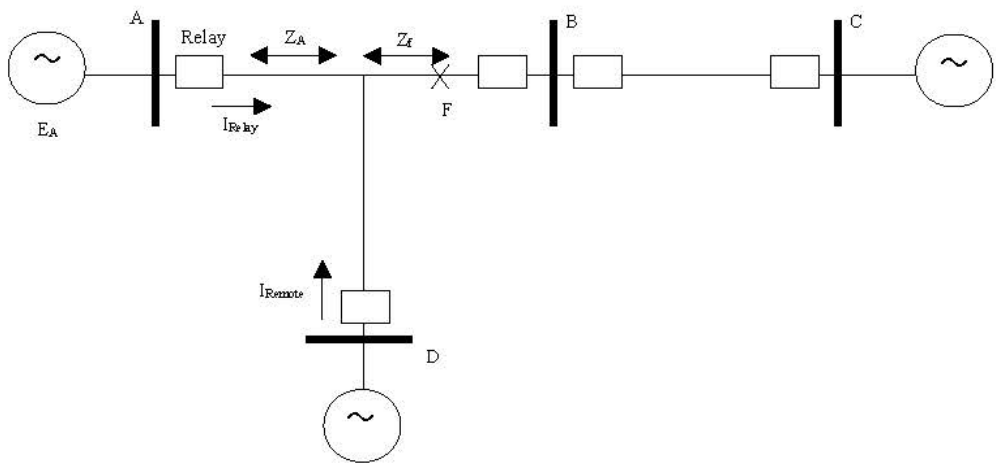

**Fig.2.** Effect of primary relays and fault locations on determining zones setting.

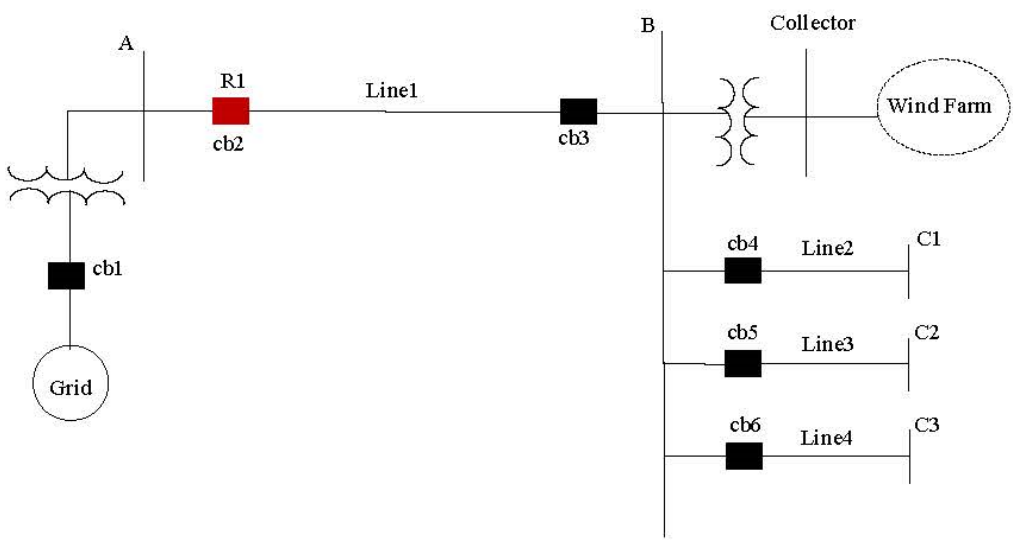

**Fig.3.** Part of the system under study

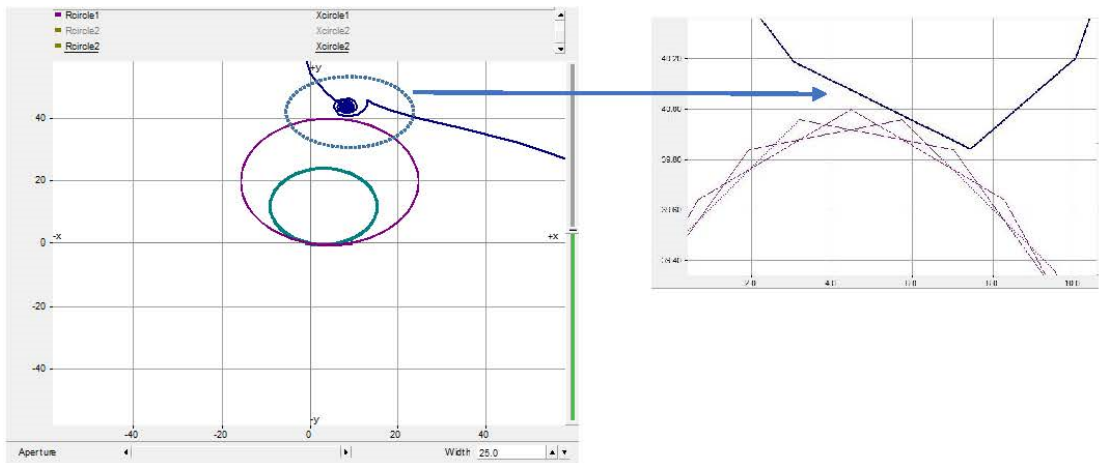

**Fig.4.** Seen impedance with no wind farm penetration during single phase to ground fault

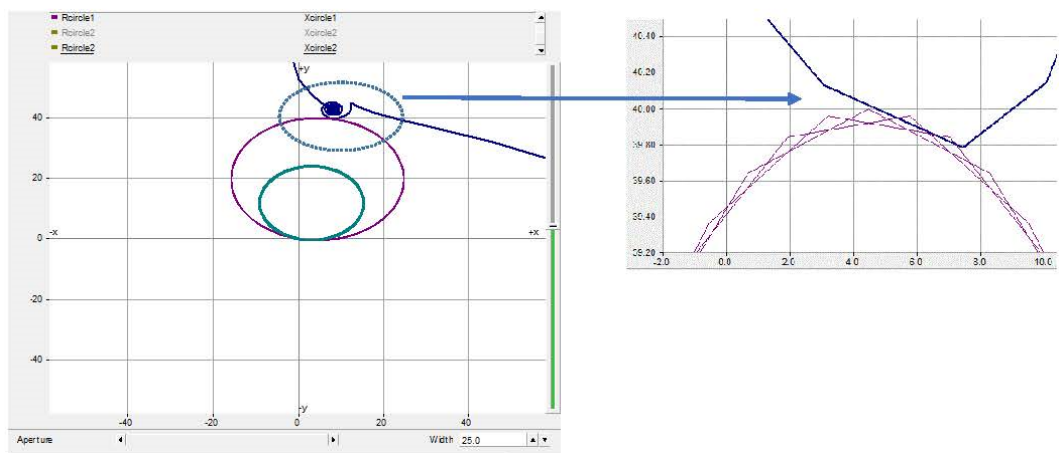

**Fig.5.** Seen impedance at 4m∕s during single phase to ground fault

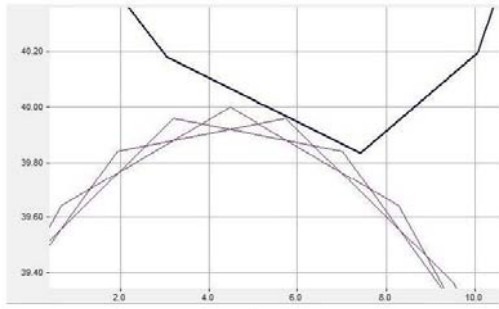

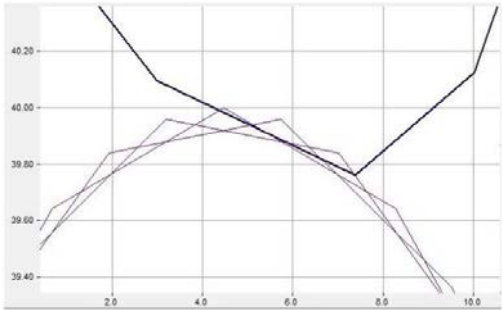

**Fig.6.** Seen impedance during single phase to ground fault (a) $16^m/_s$ wind speed (b) $22^m/_s$ wind speed

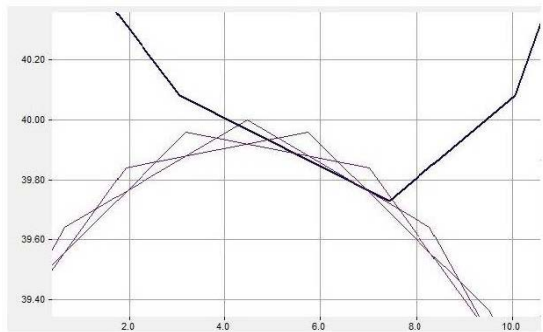

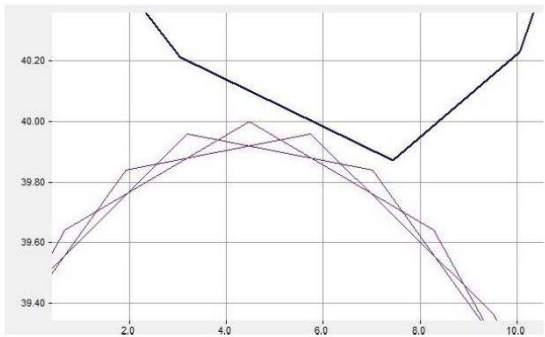

**Fig.7.** Seen impedance during single phase to ground fault (a) $17^m/_s$ wind speed (b) $23^m/_s$ wind speed

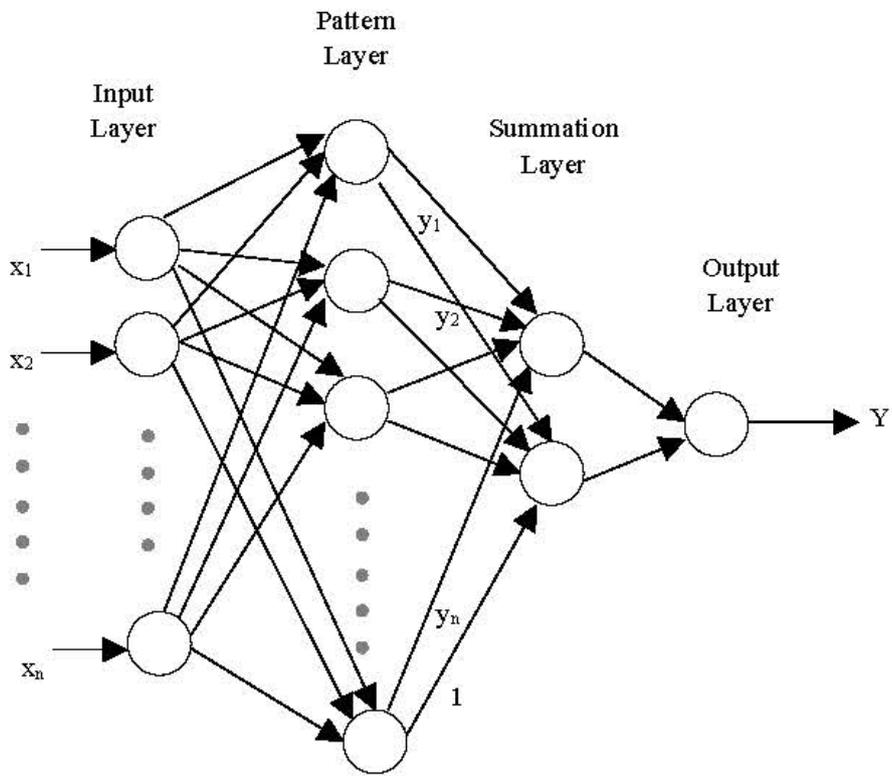

**Fig.8.** Feed Forward Network with sigmoid hidden and linear output neurons

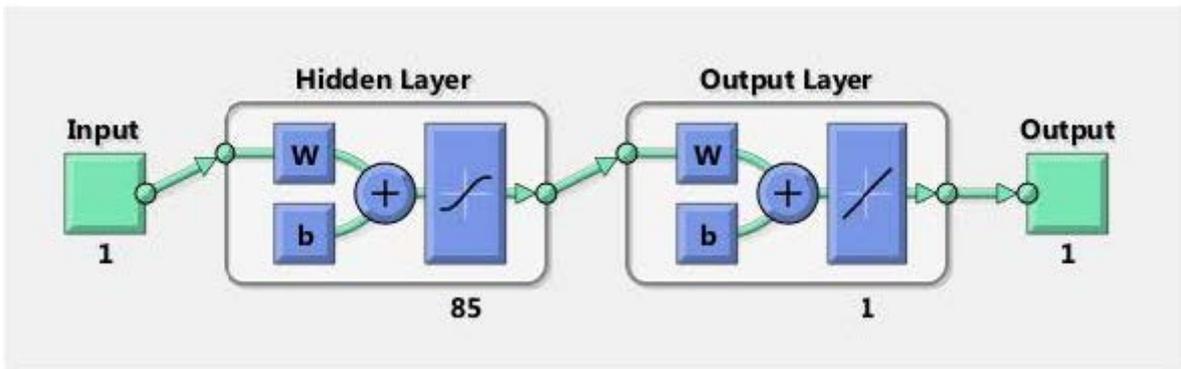

**Fig.9.** A block diagram of the Neural Network used in this study

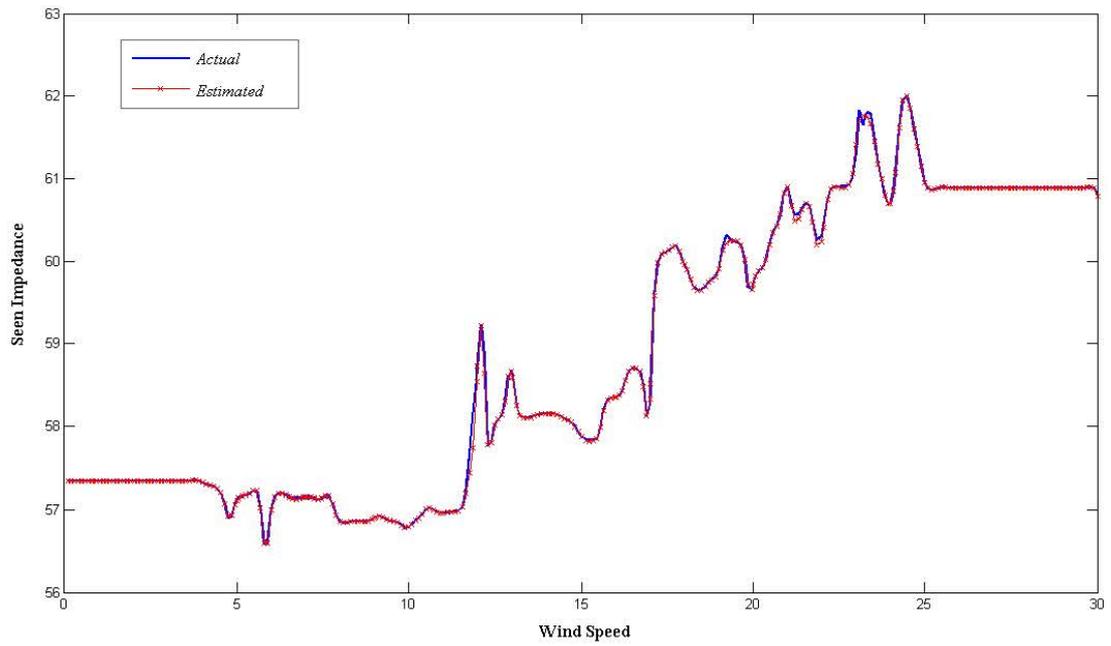

**Fig.10.** Estimated impedances characteristic obtained from neural network output compare to the actual impedances

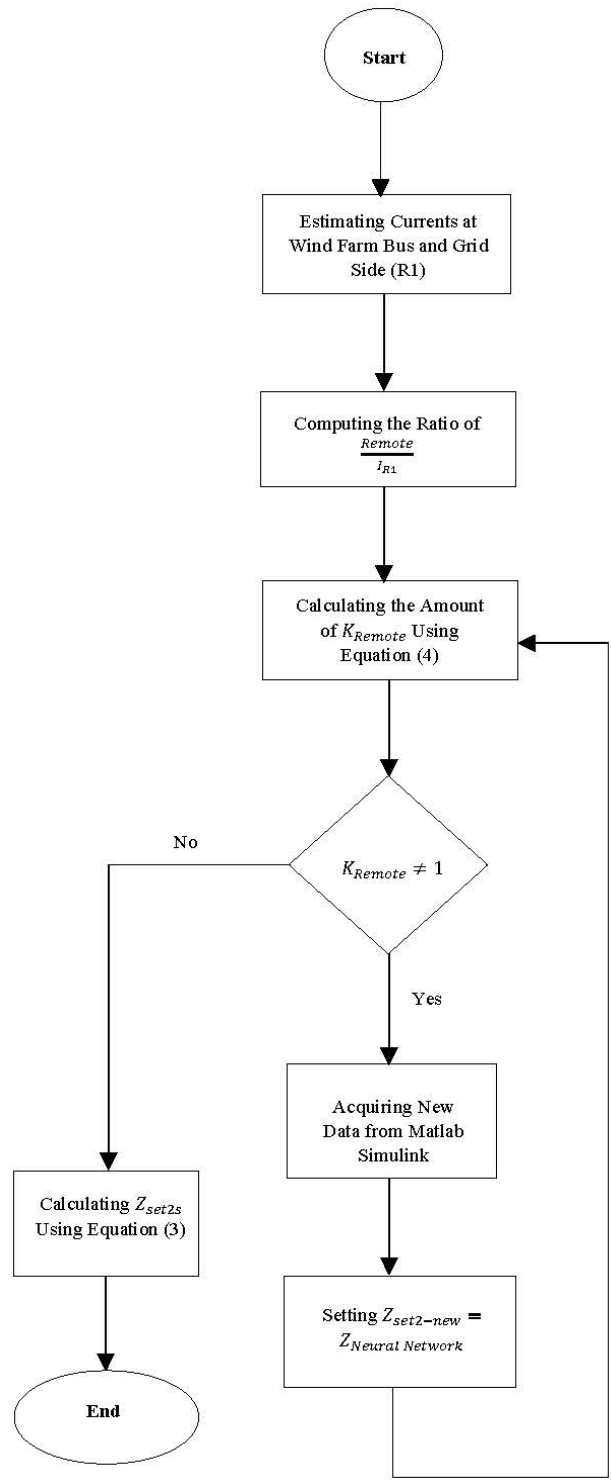

**Fig.11.** Flowchart of the proposed adaptive scheme